\documentclass{article}
\usepackage{amsmath,amssymb,setspace}
\usepackage{slashed}
\usepackage{graphicx}
\usepackage{url}
\usepackage{hyperref}
\usepackage{color}

\def\beq{\begin{eqnarray}}
\def\eeq{\end{eqnarray}}
\def\bea{\begin{eqnarray*}}
\def\eea{\end{eqnarray*}}

\def\centeron#1#2{{\setbox0=\hbox{#1}\setbox1=\hbox{#2}\ifdim
\wd1>\wd0\kern.5\wd1\kern-.5\wd0\fi
\copy0\kern-.5\wd0\kern-.5\wd1\copy1\ifdim\wd0>\wd1
\kern.5\wd0\kern-.5\wd1\fi}}
\def\ltap{\;\centeron{\raise.35ex\hbox{$<$}}{\lower.65ex\hbox{$\sim$}}\;}
\def\gtap{\;\centeron{\raise.35ex\hbox{$>$}}{\lower.65ex\hbox{$\sim$}}\;}

\newcommand{\newc}{\newcommand}
\newc{\qbar}{{\overline q}}
\newc{\Kahler}{K\"ahler }
\newc{\deltaGS}{\delta_{\rm GS}}
\newc{\QQ}{{~\bf{\color{blue}*****}}~}

\begin{document}
\begin{titlepage}
\begin{flushright}
{\large SCIPP 13/05\\
}
\end{flushright}

\vskip 1.2cm

\begin{center}

{\LARGE\bf{Moduli or Not}}

\vskip 1.4cm

{\large  Milton Bose, Michael Dine, and Patrick Draper}
\\
\vskip 0.4cm
{\it Santa Cruz Institute for Particle Physics and
\\ Department of Physics,
     Santa Cruz CA 95064  } \\
\vskip 4pt

\vskip 1.5cm

\begin{abstract}
Supersymmetry and string theory suggest the existence of light moduli.  Their presence, or absence, controls the realization of supersymmetry
at low energies.
If there are no such fields, or if all such fields are fixed in a supersymmetric fashion, the
conventional thermal production of LSP dark matter is possible, as is an anomaly-mediated (``mini-split") spectrum.   On the other hand, the axion
solution to the strong CP problem is not operative, and slow roll inflation appears difficult to implement. If there are light moduli, a mini-split spectrum is less generic, WIMP dark matter appears atypical, and the supersymmetry scale is likely tens of TeV or higher.

\end{abstract}
\end{center}

\vskip 1.0 cm

\end{titlepage}
\setcounter{footnote}{0} \setcounter{page}{2}
\setcounter{section}{0} \setcounter{subsection}{0}
\setcounter{subsubsection}{0}

\section{Introduction}

Both the relatively large mass of the Standard Model-like Higgs boson discovered at the LHC~\cite{:2012gk,:2012gu} and current bounds on superpartners place tension on models of weak-scale supersymmetry. It is possible to achieve a Higgs mass of 125 GeV in non-minimal models while placing superpartners just beyond the reach of current searches and minimizing the usual measures of fine-tuning~\cite{Hall:2011aa}. However, within the minimal model, a Standard Model-like Higgs boson at $125$ GeV is suggestive of SUSY-breaking in the range
of $10$s of TeV or higher~\cite{Giudice:2011cg,AGKWZ,ACDV,acharyakane,Draper:2011aa,yanagidasupergravity}, corresponding in the most naive estimate to parameter tuning at a level of part-in-$10^4$ if the cutoff scale is high. Thus it is worth considering the possibility that if supersymmetry plays a role in nature, the scale of its
breaking is higher than expected from conventional ideas of naturalness.

Prior to the LHC exclusions and the Higgs discovery, at least two arguments pointed to a SUSY-breaking scale above (likely
well above) $10$ TeV. The first is the experimental constraints on flavor-changing neutral currents and CP-violation. If phases and mixings are $\mathcal{O}(1)$, these processes already probe SUSY scales in the range of hundreds of TeV (see, for example, the recent study of~\cite{McKeen:2013dma}.) However, if phases and mixings are small, either accidentally or due to some structure, the bounds are diluted. 

The second argument arises from the cosmological moduli problem~\cite{bkn,nucleosynthesisbound}. Moduli (strictly speaking pseudomoduli) are
ubiquitous in string compactifications.   They are characterized by the property that sufficiently far
away in the field space, the potential for these fields vanishes.   In theories of low scale supersymmetry breaking,
one expects the masses of these fields to be of order the gravitino mass or larger.  While on the one hand such fields might seem problematic,
on the other they might play al role in understanding two pressing problems in particle physics and cosmology:  the strong CP problem
and inflation.  Many string moduli respect discrete shift symmetries, which have the potential to give rise to accidental, continuous Peccei-Quinn
symmetries.  Moreover, fields with very flat potentials would seem desirable to account for slow roll inflation.
But quite generally, if present, moduli have (other) profound consequences for cosmology.  Except under special circumstances,
before these fields settle into their ground state, for a long period they dominate the energy density
of the universe.  If the moduli have Planck-suppressed couplings, then unless they are quite heavy, they decay long after nucleosynthesis, destroying light elements and spoiling the success of this pillar of the Big Bang theory.  Moduli masses must be
at least $10$s of TeV and probably larger if their lifetimes are to be sufficiently short.  In this case, however, LSP dark matter is not produced thermally.  The possibility
that a stable LSP might be non-thermally produced via moduli decays has been widely discussed (see, for example,~\cite{moroirandall,moroirecent}.)


Therefore several considerations
point to a surprisingly high scale of supersymmetry breaking.
Adopting this ``heavy SUSY" viewpoint raises questions:
\begin{enumerate}
\item  Even if supersymmetry is broken at a high scale, might some states remain parametrically lighter and be directly visible at the LHC? Are the heavy states indirectly visible in
experiments searching for electric dipole moments or rare decays?
\item  Once one has admitted some degree of fine tuning, how much fine tuning might there be?  Are there any upper bounds on the supersymmetry breaking scale, following (for example) from cosmological
considerations or coupling unification?
\end{enumerate}

One interesting suggestion as to how LHC-observable phenomena might emerge from supersymmetry
breaking at a very high scale has been dubbed ``mini-split supersymmetry"~\cite{split,AGKWZ,ACDV}.  In these models, gauginos are significantly lighter than
other superpartners, typically with a wino-like LSP.   It is possible in such cases that the gauginos may be seen at the LHC, even though the scale of the other superpartners is much higher.
More generally, several arguments
for a split structure have been advanced:
\begin{enumerate}
\item  Because of symmetries, the gaugino/sfermion mass hierarchy is generic (it is probably not meaningful to say this is ``natural", as the structure requires significant fine tuning.) 
\item  The split spectrum is compatible with unification, perhaps even more compatible than more ``natural" models.
\item  The lightest gaugino yields a dark matter candidate.
\end{enumerate}
These features are present in models of anomaly-mediated SUSY-breaking~\cite{Giudice:1998xp,Randall:1998uk,Ibe:2006de,Ibe:2011aa,yanagida1},
as well as in attempts to build string models based on $G_2$ manifolds~\cite{acharyakane} and on variations of
the KKLT scenario~\cite{lindedudas}.

In this paper, we point out that the existence of moduli, and their nature, is a controlling issue in the realization of supersymmetry:  the breaking scale of 
supersymmetry, possible hierarchies of supersymmetric particles, and the nature of and production of dark matter are governed by the presence or
absence of moduli.  In this framework, our main focus will be
gravity mediation, though we will make some comments on gauge mediation (see also~\cite{Fan:2011ua}.)  

Our principle observation is that one can enumerate {\it three} possibilities for moduli in supersymmetric theories, each with distinct consequences for low-energy physics:  
\begin{enumerate}
\item  No moduli:  In this case, dark matter can be produced thermally.  Split supersymmetry is a likely outcome, but there
is no compelling setting either for inflation or the resolution of the strong CP problem.  A variant is the possibility that all moduli
are charged under an unbroken (or nearly unbroken) symmetry~\cite{dinemoduli}.
\item  Supersymmetric moduli (only):  Here (all of) the moduli $\rho$ have $\vert F_\rho\vert \ll m_{3/2} M_p.$  Thermal production of dark matter requires extremely heavy moduli; non-thermal
production in
moduli decays requires a slightly lower scale.   A split spectrum is
likely, but the anomaly-mediated contributions are not necessarily dominant.  Again, there is no
attractive setting for the Peccei-Quinn solution of the strong CP problem, but supersymmetric moduli are candidate inflatons.  
\item  Non-supersymmetric moduli:  an anomaly-mediated spectrum appears non-generic.  If a stable LSP is kinematically accessible, it is overproduced.  The dark matter, then,
needs to be something other than WIMPs, such as axions.  For fixed axion decay constant,  $f_a$, there is an upper limit on the modulus mass. 
\end{enumerate}
The first two possibilities do not violate any obvious principle of theories of gravity, but string theory examples of these phenomena are hard to come by. The
third scenario requires moduli masses  in the $100$ TeV range, and was initially viewed with skepticism because of the resulting fine tuning.
But given the lack of compelling examples of the first two solutions,
the possibility of heavy non-supersymmetric moduli and correspondingly high scale of supersymmetry breaking has always demanded serious attention. 

This paper will elaborate these points.  In Sections~\ref{nomoduli},~\ref{Smoduli}, and~\ref{NSmoduli} we discuss the cases of no moduli, supersymmetric moduli, and non-supersymmetric moduli. In Section~\ref{axions} we discuss constraints on moduli if dark matter is composed of axions and in particular upper bounds
on the moduli masses, and in Section~\ref{strongmoduli} we discuss briefly the case of scalar fields with stronger-than-Planck-strength interactions\cite{nemeschansky,
dinemoduli,lindedudas}. In Section~\ref{conclusions} we conclude, summarizing how the moduli scenarios we have outlined
 control the nature and phenomenology of low-energy supersymmetry.

\section{Absence of Moduli}
\label{nomoduli}

The moduli observed in string models might be artifacts of theorists' efforts
to construct weak coupling quantum gravity theories.  It is possible that 
nature may exhibit low energy supersymmetry without moduli.  Supersymmetry may be broken in a sector of the theory without
gauge singlet fields with large $F$ components.  This is typical of many models of dynamical supersymmetry
breaking such as the ISS models with metastable supersymmetry breaking and models with stable dynamical supersymmetry breaking.  In these models one would expect the leading contribution to gaugino masses to arise through anomaly mediation, leading to a split spectrum. Refs.~\cite{Nelson:2002sa,Hsieh:2006ig,AGKWZ} have stressed that  certain threshold effects can lead to a spectrum which is not strictly anomaly mediated.
In the next section, we will see another way in which a more ``compressed" spectrum might readily arise.
 
In a theory without moduli, if the universe was already in thermal equilibrium at very high energies, a viable thermal relic abundance of a wino-like LSP may be produced.  As noted in~\cite{Hisano:2006nn,moroirecent}, the wino mass in this case tends to be large (2.7-3 TeV), and, with a strictly anomaly-mediated spectrum, all of the gauginos are far beyond the reach of the LHC.

While in some respects very simple, the no-modulus scenario has unappealing features.   First, as we have noted, moduli would seem likely
candidates for axions, and this possibility is unavailable.  As we will describe in Section \ref{strongmoduli}, a more complicated (and perhaps less
plausible) structure is necessary to implement the Peccei-Quinn solution of the strong CP problem in such theories.
Second, from the point
of view of slow roll inflation, the absence of moduli is troubling.  One could certainly imagine that the role of the inflation is played by a field with a potential which is flat in some
suitable region of field space, but moduli appear ready-made to satisfy the conditions for slow roll inflation.

\section{Supersymmetric Moduli}
\label{Smoduli}

By a supersymmetric modulus, we mean a modulus with a mass parametrically larger than $m_{3/2}$.  Because the mass is supersymmetry
preserving, it should arise from a mass term in the superpotential, while in order that the field be considered a modulus, its higher order couplings
must be small.  
We can parametrize the superpotential as
\beq
W_\phi = m_\phi M_p^2 w(\phi/M_p).
\label{wofs}
\eeq
Perhaps the most well-known model containing supersymmetric moduli is the KKLT scenario, for which the superpotential has
this form, as we will review shortly.  First let us consider a simple toy model. We can define the origin for $\phi$ so that $W$ contains no linear term in $\phi$. 
To determine the typical size of $\langle \phi \rangle$ and $\langle F_\phi \rangle$, we need to include supersymmetry-breaking dynamics.  Suppose $W$ has the form of Eq.~(\ref{wofs}), and an additional piece responsible for supersymmetry breaking,
\beq
W=&W_\phi +W_0+f X\;.
\eeq
We suppose that the~\Kahler potential is such that $X$ is stabilized at the origin.  Then including a general~\Kahler potential
for $\phi$, 
\beq
K=& (k^\phi_1\phi+c.c.)+ \phi^\dagger\phi+(k^\phi_3 \phi^\dagger \phi \phi+c.c.)\;.
\eeq
At the minimum,
\beq
\phi \simeq k_1^\phi {m_{3/2} \over m_\phi}M_p \;,
\eeq
and the $F$ component of $\phi$ is of order
\beq
F_\phi \simeq k_1^\phi {m_{3/2} \over m_\phi} m_{3/2} M_p\;.
\eeq
In such models, $\phi$ can couple to $W_\alpha^2$ with $\mathcal{O}(1)$ coefficient and maintain the minimal anomaly-mediated spectrum as long as $m_\phi$ is at least two or three orders of magnitude larger than $m_{3/2}$.

\subsubsection{KKLT}

The scenario popularized by KKLT provides a model for supersymmetric moduli of the type we have described.  It also illustrates possible additional
problems with such cosmologies.  The model is described by an effective Lagrangian for a field, $\rho$,
with superpotential:
\beq
W = e^{-b \rho} + W_0\eeq
with small $W_0$.  The~\Kahler potential is:
\beq
K = - \ln(\rho + \rho^\dagger).
\eeq
The model has a supersymmetric minimum with
\beq
\rho \approx {1 \over b} \log(W_0/b).
\eeq
At the minimum, $\rho$ is large.
Supersymmetry must be broken by some other dynamics.  It is often argued that there can be explicit breaking by $D$ branes, but it is not clear that this
is consistent.  A simple possibility is that there are some other light degrees of freedom which spontaneously break supersymmetry~\cite{dineintermediate,nillesfterms,lindedudas}.
For example, introduce a field $X$ with superpotential
\beq
W_X = f~X
\eeq
and a~\Kahler potential
\beq
K_X = a X + {\rm c.c.} + X^\dagger X + \dots
\eeq 
where the higher order terms are chosen so that $X=0$ at the minimum of the potential (this is a definition of the zero of $X$).
Then we can relate $m_{\rho}$ to $m_{3/2}$:
\beq
m_\rho^2 = \rho^2 m_{3/2}^2
\eeq
Supersymmetry breaking induces a shift in $\rho$ of order
\beq
\delta \rho \sim {1 \over \rho}
\eeq
and a corresponding shift in $F_\rho$.  
$F_\rho$ is suppressed relative to $m_{3/2} M_p$.  In particular,
\beq
e^K \vert F_\rho \vert^2 g^{\rho \rho^\dagger} \sim m_{3/2}^2 M_p^2 \left ({m_{3/2} \over m_\rho} \right )^2.
\eeq
If we suppose that $m_{3/2} \approx 10 ~{\rm TeV}$, and that $\rho \sim {4 \pi  \over \alpha_{gut}}$, then 
the reheating temperature (assuming $\rho$ is the only modulus) is greater than 5 GeV, in a range such that
one can produce a suitable dark matter density.  Of course, it is critical that $\rho$ is the {\it only} light modulus; other
moduli~\cite{dineaxions,bobkovraby} breaking supersymmetry lead to cosmological difficulties.  The $X$ field above is such a modulus and would need to be replaced by sector which dynamically breaks SUSY without moduli.  As in the no-modulus case, this could be a theory with stable, dynamical supersymmetry breaking, or a theory
with metastable breaking, such as ISS\footnote{One of the scenarios discussed in \cite{lindedudas} is a realization of this latter possibility.}.
  
We might expect a $\rho W_\alpha^2$ coupling.  The non-zero $F_\rho$ will then contribute to gaugino masses.
This contribution is of order
\beq
m_\lambda \approx {m_{3/2} \over \rho}.
\eeq
The anomaly-mediated contributions then only dominate if $\rho$ is sufficiently large (or equivalently the modulus is quite heavy compared to $m_{3/2}$).

There are other cosmological issues associated with such moduli, particularly the problem of overshoot~\cite{brusteinsteinhardt} and related
destabilization issues.
Various solutions to this problem have been proposed.  Specifically in the framework of KKLT models, ``racetrack" type superpotentials \cite{Kallosh:2004yh,BlancoPillado:2005fn,Kallosh:2007wm} may naturally lead to heavy moduli which avoid these difficulties.  They are also argued to lead
to anomaly-mediated gaugino masses~\cite{Linde:2011ja,lindedudas}.  Other solutions have been discussed, for example,
in \cite{dinebrusteinsteinhardt}; as our focus is on somewhat different issues, we will not assess these scenarios further here.

One can contemplate variants of the scenario where the would-be modulus acquires mass comparable to the Planck mass
~\cite{dineintermediate}\footnote{KKLT presumes that the superpotential for the modulus contains a small constant, $W_0$.  It is conceivable that this constant
is large, and that the effective low energy theory, {\it after} integrating out this modulus, has a small $\langle W \rangle$, required for
a small cosmological constant.  Under these circumstances, the modulus could be quite heavy.}.  This would be 
a realization of the no-moduli scenario (in the absence of a pseudomodulus, i.e. replacing $X$ by a model of dynamical supersymmetry
breaking without moduli.)

\subsection{Consequences of Supersymmetry for Moduli Decays}
\label{susydecayrates}

It is straightforward to show that the decay rates into the scalar and fermionic components of a lighter multiplet are related in specific ways by supersymmetry, up to corrections proportional to the soft masses. In Appendix A we discuss how this works at tree level for the dimension 5 operators mentioned previously. In Appendix B we outline a more general argument from the unitary representations of the SUSY algebra. Here, for brevity, we sketch an argument from field theory for the case of decays to a massless multiplet.  Consider first a simple Wess-Zumino model with a heavy field, $\Phi$, and a massless field, $\phi$.  For the superpotential, take
\beq
W = {1 \over 2} M \Phi^2 + \lambda \Phi \phi \phi.
\eeq
Supersymmetry relates the Green's functions:
\beq
\langle F^*_\Phi(x_1) \psi_\alpha(x_2) \psi_\beta(x_3) \rangle  
\epsilon^{\alpha \beta} =2 \langle \Phi(x_1)^* \partial_\mu \phi(x_2)\partial^\mu \phi(x_3) \rangle\;.
\label{wardidentity}
\eeq
This relation can be proven easily, for example, by 
considering the superspace Green's function:
\beq
\langle \Phi^*(x_1,\theta_1)\phi(x_2,\theta_2) \phi(x_3,\theta_3) \rangle
\eeq
The left hand side of Eq.~(\ref{wardidentity}) is the coefficient of $\bar \theta_1^2 \theta_2 \theta_3$ in this Green's function; translating by $\theta_1$ in superspace,
the coefficient of this term is the right-hand side of the equation. 

To extract the decay amplitudes, we can apply the LSZ formalism.  First we note the relations for the Green's functions, in momentum space,
\beq
\langle F^\dagger F \rangle = p^2 \langle \phi^\dagger \phi \rangle.
\eeq
So we can relate the single particle matrix elements needed for LSZ; those of $\phi$ and $F$ differ by a factor of $m^2$, the physical
on-shell mass.  There are two possible initial states (which can be thought of as the scalar and its antiparticle) and two possible final states in either the
two boson or two fermion channel.  Combining the Ward identity for the Green's functions and the result for the single particle matrix elements demonstrates the
equality of the two boson and two fermion matrix elements.  The result is readily verified at tree level.

Similarly, for a scalar coupled to $W_\alpha^2$, one can prove an equality for the matrix elements (and hence the rates) for the decays: $\phi \rightarrow A_\mu + A_\mu$ and $\phi \rightarrow \lambda \lambda$. When supersymmetry is broken these equalities will fail, but, except for tuned values of the
parameters, we expect the rates to be comparable.

\subsection{Moduli Decays and the Reheat Temperature}

We can consider, then, the lifetime of the moduli (first in the supersymetric case).  The lifetimes depend on
the kinetic terms for the moduli and their couplings to other fields and one can obtain
quite different results with different choices.  Given that much of the motivation to consider moduli comes
from string theory, it seems appropriate to consider
kinetic terms familiar from various string models. We
take as a model the heterotic string
compactified on a Calabi-Yau manifold, and take the modulus to be the so-called model-independent dilaton.
Then the \Kahler potential and gauge coupling functions are~\cite{wittendimensionalreduction}:
\beq
K = -M_p^2 \ln(S + S^\dagger);~~~f= S\;.
\eeq
Here we have taken $S$ to be dimensionless and indicated explicit factors of $M_p$.  
In this case where the decay is principally through the coupling $S W_\alpha^2$,
the decay rates to pairs of gauge bosons and gauginos are the same.  At leading order, summing over the gauge multiplets of the MSSM, one obtains~\cite{nakamurayamaguchi}
\beq
\Gamma(S \rightarrow g g)+\Gamma(S \rightarrow \tilde{g} \tilde{g}) = {3 \over 4\pi} {m_S^3 \over M_p^2}\;.
\eeq
This translates to a reheating temperature
\beq
T_R = 9.8 \times \left ( {g_*(T_R) \over 10} \right )^{-1/4} \left ({m_S \over 10^5 ~{\rm GeV}} \right )^{3/2} ~{\rm MeV}\;.
\label{trgluino}
\eeq
For reference, we note that the minimum temperature required to achieve successful nucleosynthesis is approximately
$4$ MeV~\cite{nucleosynthesisbound}.

An alternative model is provided by the ``T modulus" of simple Calabi-Yau compactifications of the heterotic string~\cite{wittendimensionalreduction}.
Here:
\beq
K = -3M_p^2 \log(T + T^* - {1 \over 3}{ \phi_i^* \phi_i \over M_p^2});~~~f=0\;,
\eeq
where the $\phi_i$ denote the matter fields.  Writing $T = T_0 + \delta T$, after rescaling the $\delta T$ and $\phi_i$ kinetic terms to make them canonical generates the couplings:
\beq
{\cal L}_{T\phi} = {1 \over \sqrt{3}}\delta T \phi_i^* \phi_i \;.
\eeq
These are among the dimension-5 operators listed in~\cite{moroirandall}.  The decay rates to fermion and boson pairs are the same in the SUSY limit and are suppressed by $m_\phi^2/m_T^2$. When a soft mass of order $m_{3/2}$ is present for the bosonic components, there is a contribution that independent of $m_\phi$, but is still suppressed~\cite{moroirandall},
\beq
\Gamma(T \rightarrow \phi \phi) \sim {1 \over 4 \pi} \left({m_{3/2} \over m_T}\right)^4{m_T^3 \over M_p^2}\;.
\eeq

Shortly, we will be interested in the non-supersymmetric case, and in particular the possibility that the decay channels to $R$-odd particles are not accessible. In that case,
in Eq.~(\ref{trgluino}), $9.8$ is replaced by $6.9$.

\subsection{Decays and the Relic Density}

The most urgent question in moduli decays is the resulting relic density.  There is the possibility of overproduction
of LSPs, if stable, and gravitinos.  These lead to too-early matter domination, inconsistent with the observed  light element abundances.  We will focus principally in this subsection on models with a conserved $R$-parity
and a stable LSP.  We will remark at the end about the effects of $R$-parity violation, postponing more detailed analysis to a subsequent work.

We will first assume a conserved $R$-parity.  In this case LSPs are produced (possibly overproduced) in decays of the modulus.  It is also necessary to consider
modulus decays to gravitinos.  While a $10-100$ TeV gravitino is relatively short-lived, its decay products include LSPs, which may be problematic.

As demonstrated in the previous section, 
in the SUSY limit,
amplitudes for two-body decays to particles are identical to those for two-body
decays to their supersymmetric partners.  In particular, there are no helicity suppressions of decays to fermions compared to decays to bosons, as has been suggested in certain contexts\footnote{Both fermionic and bosonic decays from the $\phi Q^*Q$ operator are proportional to small supersymmetric masses in the SUSY limit; therefore the leading effect of this operator may be the $m_{3/2}^4$ contribution to the bosonic final states.}.
With heavy supersymmetric moduli, all $R$-odd decays to partners of Standard Model fields
are kinematically allowed and occur with rates approximately equal to the rates into their $R$-even partners, since the light MSSM fields appear supersymmetric to the moduli.

As a result, the number of LSPs produced per modulus decay is $\mathcal{O}(1)$.  To keep the reheating temperature above the temperature of nucleosynthesis requires
moduli masses above 30-100 TeV.  In this range, the LSP density is an $\mathcal{O}(1)$ fraction of the total energy density
at temperatures of order a few MeV, so matter domination occurs far too early. The weak interactions freeze out at 
\begin{align}
T_F\sim (M_p^{-2}G_F^{-4}T_R)^\frac{1}{7}\;,
\end{align}
which is about 1.8 MeV for $T_R\sim 5$ MeV, compared with freeze-out at 0.8 MeV in the ordinary radiation-dominated universe. The neutron-to-proton ratio thus increases from $n/p\approx 1/6$ at weak freeze-out to $n/p\approx 1/2$, increasing the abundance of helium. 

A simple solution to this
problem is that the moduli are heavier than $10^6$ GeV, producing
a reheating temperature of order a few hundred MeV or higher.  For supersymmetric moduli,, this large mass scale is not disturbing.  If the dark matter annihilates effectively,
the reheating temperature may be lower. In that sense an anomaly-mediated-type spectrum may in fact seem favored, since $\sigma_{\rm wino}\sim m_{\tilde{W}}^{-2}$.

However, a related problem may still arise for supersymmetric moduli, dubbed the ``moduli-induced gravitino problem"~\cite{endo}.   If moduli decays to gravitino pairs occur with $\mathcal{O}(1\%)$ branching fraction, the decays of gravitinos still typically overproduce dark matter, even if they avoid BBN constraints.  As pointed out
in \cite{Dine:2006ii}, exploiting the Goldstino equivalence theorem allows analysis of this problem by considering couplings of the modulus $S$
to Goldstinos. The branching fraction
to gravitinos is controlled by~\Kahler potential couplings of $S$ to the Goldstino superfield,\footnote{Here we mean in the sense
of non-linear realizations of supersymmetry, as in~\cite{ks}; we are not assuming the gravitino has a light
supersymmetric particle.} $S^\dagger Z Z + ~{\rm c.c}$.  This coupling
might be suppressed (see, for example,~\cite{Kaplan:2006vm}); if not, the branching ratio of the modulus to gravitinos is of order one.  So whether this is a problem
depends on microscopic details of the theory.

We note in passing that with supersymmetric moduli, baryons might be produced coherently or in decays of the modulus.  If there are $\epsilon$ baryons produced per modulus, the baryon to photon ratio
is 
\beq
{n_B \over n_\gamma} \approx \epsilon {T \over m_\phi} \approx \epsilon \left ( {T \over M_p} \right ) ^{1/3}.
\eeq
So, for example, for a $1$ GeV reheating temperature, we require $\epsilon \approx 10^{-4}$.   Alternatively, if $\epsilon$ is
fixed by the microscopic theory, the mass of the modulus is determined.

So far, we have assumed a conserved $R$-parity.  If $R$-parity is violated, the role of the dark matter must be played by some
other field.  Provided the $R$-parity violating couplings are not too small, the lifetime of the would-be LSP is much shorter than that
of the moduli, so their production is not a cosmological issue.  For example, if the principle source of $R$ breaking is the coupling
\beq
W_R = \lambda \bar t \bar b \bar s
\eeq
then, unless $\lambda < 10^{-10}$ or so, gaugino decays are sufficiently rapid.

\subsection{Summary}

Supersymmetric moduli are a plausible outcome of moduli-fixing.
They are suggestive of a split spectrum for superparticles, though anomaly-mediated
contributions do not necessarily dominate the gaugino spectrum.   In such cases, avoiding overproduction of LSPs
sets a lower bound on the modulus mass.  Avoiding overproduction of dark matter through gravitinos
places restrictions on the microscopic details of SUSY-breaking.

\section{Non-Supersymmetric Moduli}
\label{NSmoduli}

The KKLT scenario, with supersymmetry broken in a sector of the theory without flat directions, provides a model for supersymmetric
moduli fixing.  But there are a number of reasons to suspect that there should be moduli which gain mass only through supersymmetry breaking effects.  The need for an axion to solve the strong CP problem provides one motivation; in a supersymmetric context, a non-supersymmetric modulus seems to provide,
as we have said, an ideal axion candidate.  A second motivation is provided by metastable dynamical supersymmetry
breaking, and especially the retrofitted models~\cite{dfs,bosedine}, where such moduli are an integral part of supersymmetry breaking.
Inflation is also suggestive
of relatively light moduli.   Successful inflation requires a mass small compared to the Hubble constant during inflation.
A non-supersymmetric modulus automatically has mass of order the Hubble constant, so only a modest coincidence is required.
Such a modulus also has only small self-interactions, so the mass can readily remain small throughout and the potential can be adequately flat.

\subsection{$F$-terms}

It is often assumed that there is only one modulus with an $F$-term of order $m_{3/2}M_p$. However, in a gravity-mediated theory, {\it all moduli} with masses 
of order $m_{3/2}$  will tend to have
$F$ components of this order, whether or not they appear explicitly
in the superpotential.   In supergravity, the $F$ component of a modulus $\phi$ is
\beq
F_\phi =g^{-1} e^{K/2} \left ({\partial W \over \partial \phi} + {\partial K \over \partial \phi} W \right ).
\eeq
Here $K$ is the~\Kahler potential and $g$ is the inverse metric on the field space\footnote{This is schematic; in the presence of multiple fields,
one needs to consider diagonalization of $g$.}.
If $\phi$ does not appear in the superpotential, in the absence of symmetries, the second term is of order $m_{3/2} M_p$ from the linear term in $K$.

Of course, it is possible by a redefinition of the fields to simply {\it define} one field to have a non-vanishing auxiliary component, while all others vanish.  But the redefinition will affect the couplings of these fields.  For example, if originally only one field couples to $W_\alpha^2$, then in general,
all will, including the linear combination with non-vanishing $F$ component.  
In this subsection, we will discuss in more detail the scaling of the moduli~\Kahler potentials and their implications for the spectrum.

In the retrofitted models, there is a modulus $X$ coupled to $W_\alpha^2$ of some new gauge group.   In the simplest
case, this hidden sector is a pure gauge theory, and $X$ does not couple directly to other light fields transforming
under this group. Gaugino condensation
in this group gives rise to a superpotential for $X$, and the $X$ vev is fixed by the~\Kahler potential.  $X$ can be defined so that
the first derivative of the~\Kahler potential, $K_1$, vanishes.  

Given that $X$ couples to the kinetic term of one gauge group, it is likely to couple to the Standard Model gauge groups as well. As is typical
of moduli of string theory, if we take the modulus to be dimensionless, its imaginary part is periodic with a period we can
take to be a multiple of $2\pi$.
The gaugino masses then depend on the gauge coupling, $g^2$, and
$K_2$, the second derivative of the~\Kahler potential at the minimum, as
\beq
m_{\lambda} \sim {g^2 \over \sqrt{K_2}} m_{3/2}.
\label{ksuppression}
\eeq
The gaugino mass can be small if $K_2$ is large, or if the $XW_\alpha^2$ coupling is for some reason suppressed.

It would seem that we are free to hypothesize whatever form for the~\Kahler potential we wish, but string theory
provides some guidance.  Typical~\Kahler potentials, as exemplified by the dilaton of the heterotic string or Type II theories, or the radial dilaton of each, behave like
\beq
K \sim -\ln(X + X^\dagger)
\eeq
where the corresponding field obeys the periodicity property (with a suitable normalization)
\beq
X \rightarrow X + 2 \pi i.
\eeq
Because of the periodicity, $X$ couples linearly to $W_\alpha^2$.
If there is a single field with such a coupling, 
\beq
\langle X \rangle = g^{-2}.
\eeq
Then $K_2$ is small and the gaugino mass is of order the gravitino mass.

With multiple fields, there are additional possibilities allowing for hierarchies between gaugino and
scalar masses.  For example, with two fields, $X_1$ and $X_2$, with $X_1 \gg X_2 \gg 1$
and $F_{X_1} \ll F_{X_2}$, then
\beq
m_\lambda = c~ m_{3/2} {1 \over X_1} \approx {g^2} m_{3/2}.
\eeq
An argument for moduli vevs of this sort appears in~\cite{wittenheterotic}.
What appears typical is that in the presence of non-supersymmetric moduli, most soft SUSY-breaking masses will be of order $m_{3/2}$, without a large hierarchy.

\subsection{Decays and the Relic Density}

In order that non-supersymmetric moduli decay before nucleosynthesis (implying a reheat temperature greater than about $10$ MeV), they should
decay through dimension 5 operators; if they decay through dimension 6, their masses need to be of order $10^7$ TeV or more.  Possible
dimension 5 operators are listed in~\cite{moroirandall} and include the aforementioned coupling to $W_\alpha^2$ as well as~\Kahler couplings to $Q^*Q$ and $H_uH_d$.

If there is a conserved $R$-parity, and a modulus can decay to the LSP, then the number of LSPs produced in a single modulus decay ($N_{LSP}$) is an important parameter controlling the cosmology.
We have already shown in Section~\ref{susydecayrates} that
in the case of unbroken supersymmetry, the decays to pairs of particles and their supersymmetric partners occur at equal rates.
For broken supersymmetry, these relations are corrected, but we do not expect qualitatively significant changes, except for kinematic reasons
in particular regions of parameter space. Consequently, we expect $N_{LSP}$ is typically $\sim 1$.  In this situation, an acceptable cosmology
only emerges for extremely heavy moduli.

\begin{figure}[!t]
\begin{center}
\includegraphics[width=0.5\textwidth]{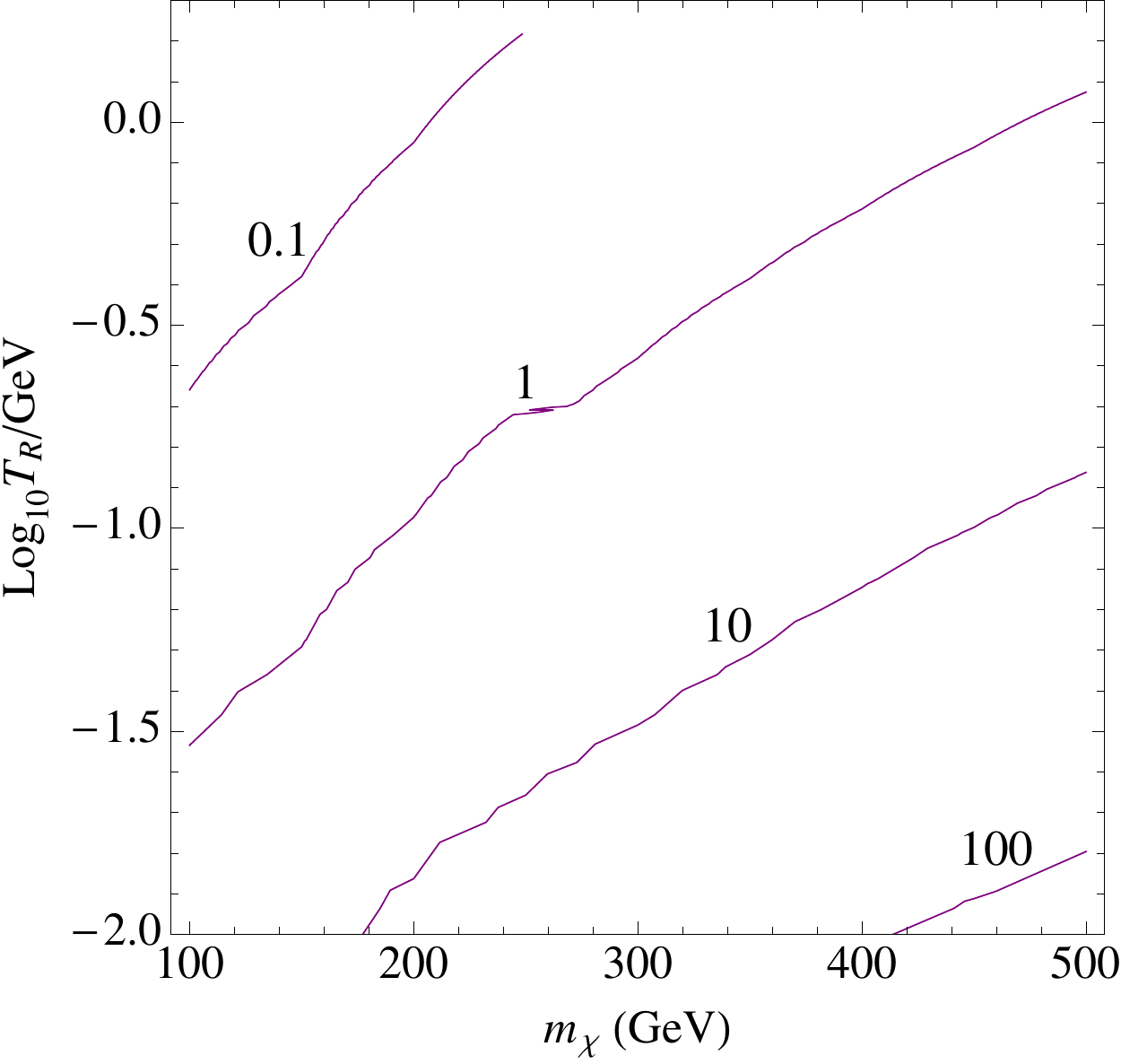}\hspace{3mm}
\caption{$\Omega h^2$ for the case $N_{LSP}=1$.}
\label{fig:nlsp1}
\end{center}
\end{figure}

When the temperature is around $10$ MeV, a large fraction
of the energy density is in LSPs. The relic density of a wino LSP for can be estimated by integrating simple Boltzmann equations~\cite{moroirandall}.  In Fig.~\ref{fig:nlsp1} we plot contours of the wino relic density as a function of the wino mass and the reheating temperature $T_R$, obtained by numerically integrating the Boltzmann equations. We fix $N_{LSP}=1$ and take 
\begin{align}
\Gamma_\phi\sim \frac{1}{2\pi}\frac{m_\phi^3}{M_p^{2}}\;,
\label{eq:gammaphi}
\end{align}
as an estimate of the total width (compatible with our earlier discussion.)

It is clear that in this scenario the reheating temperature should be over a few hundred MeV, corresponding to a SUSY-breaking scale above $10^6$ GeV, and the wino should be extremely light compared to $m_{3/2}$, so that pair annihilation is more effective at reducing the density (a thermal abundance of $\Omega h^2\sim0.1$ is not achieved until the reheating temperature is of order several hundred GeV.) We have already argued that such a spectrum is atypical in the presence of non-supersymmetric moduli. 

Similar results have been obtained recently in~\cite{moroirecent}, where it was also found that a Sommerfeld enhancement can greatly increase the wino pair annihilation rate and suppress the relic density to acceptable values for any reheating temperature. However, this effect is only present in a narrow window around $m_{\tilde{W}}\sim2.4$ TeV and is thus highly non-generic. The authors of this paper also stressed the clear degeneracy in Eq.~(\ref{eq:gammaphi}): if the width is increased by decreasing the effective cutoff scale, the reheating temperature can be set high enough for any value of the modulus mass. We will discuss the possibility of modulus couplings stronger than expected with Planck suppression in Section~\ref{strongmoduli}.

To suppress the relic abundance without going to extreme regions of parameter space, the most obvious possibility is to make the moduli lighter
than the LSP.  An alternative possibility is that there is no conserved $R$-parity and no stable LSP.  In either case, the universe
might simply ``reheat" to temperatures above nucleosynthesis, without leaving stable (or cosmologically long-lived) relics.

Another possibility is that the gauginos are light and modulus decays to these lighter states are somehow suppressed.
For example, it might be that all couplings of the non-SUSY moduli to $W_\alpha^2$ vanish or are highly suppressed (by a large value of $K_2$ in
Eq.~(\ref{ksuppression}).)
Then the 
gauginos are indeed parametrically lighter than other superparticles. It is necessary that the moduli decay through other dimension-5 operators, such as $\phi Q^*Q$ and $\phi^*H_uH_d$. If the Higgsinos and sfermions are heavier than $m_\phi$, the width to gaugino final states {\it will} be suppressed; to keep the total width of order Eq.~(\ref{eq:gammaphi}), decays to the Higgs bosons must be unsuppressed. The latter condition may be satisfied if Higgsino masses of order $m_{3/2}$ are obtained from an order-one $\phi^*H_uH_d$ coupling. Such a scenario can perhaps lead to an acceptable 
dark matter density, though the Higgsinos and sfermions may have to be rather heavier than the moduli.  As we have remarked, this sort of spectrum
seems surprising from the perspective of known string models, but it is a logical possibility.

We note that it has recently been pointed out in~\cite{decaystoaxions} that there is another cosmological problem with moduli that are stabilized by SUSY-breaking dynamics. The axion component of the modulus multiplet may remain very light and generically the branching ratio of moduli into axion pairs will be sizable. This scenario is constrained by the Planck measurement of $\Delta N_{\rm eff}$ and the authors of~\cite{decaystoaxions} emphasize that simply raising $m_{3/2}$ is insufficient to evade the bounds.  The constraints imposed on particular types of microscopic models by this phenomenon will be described elsewhere.

\subsection{Summary}

The $F$-terms of non-supersymmetric moduli are typically of order $m_{3/2}M_p$, which implies that the generic spectrum is not split.  If $R$-parity is conserved and gaugino masses are comparable to those
of other superparticles, viable cosmology demands kinematic suppression of the decays of moduli to the LSP. The simplest possibility
is that any modulus in the theory must be lighter than the LSP.  Alternatively, gauginos might be light, which requires that moduli couplings to $W_\alpha^2$ must vanish or be quite small. If there is no branching ratio suppression, the reheat temperature should be quite high in order to avoid overclosure, corresponding to a high SUSY-breaking scale. Sufficient branching ratio suppression can be achieved for lower $m_{3/2}$ if decays to sfermions and higgsinos are kinematically forbidden and the coupling to $W_\alpha^2$ is small (as it should be to keep the gauginos light.)  As pointed out in \cite{decaystoaxions}, decays of non-supersymmetric
moduli to axion-like objects place significant constraints on the microscopic theory, but suggest that the effective number of neutrinos at nucleosynthesis 
may be larger than three.

\section{Axions as Dark Matter, Baryogenesis, and Upper Bounds on $m_{3/2}$}
\label{axions}

Except for models with extremely heavy moduli, or no moduli at all, we have seen that it is challenging for the LSP to be the dark matter.
An alternative dark matter candidate is the axion.  In this section, we explore this possibility, discovering that for a fixed axion decay constant (or more
precisely, $\theta_0 f_a$, where $\theta_0$ is the initial axion ``misalignment angle"), there is an {\it upper} bound on the mass of the modulus.
We will also consider in this section the question of baryogenesis.  Again, given the low reheat temperature, there appear to be two
possibilities:  baryon number violation in the moduli decays, and Affleck-Dine baryogenesis~\cite{ad}.

We have stressed that the ``no modulus" or ``all moduli heavy" scenarios are unlikely settings for the axion solution to the strong CP 
problem, since if supersymmetry survives to the multi-TeV scale, the would-be modulus partner of the axion
is missing.  In such theories, one would need to introduce a Peccei-Quinn symmetry along the lines we will discuss in Section \ref{strongmoduli}.

Therefore, we assume the existence of moduli with masses of order $m_{3/2}$, and an axion to solve the strong CP problem.
We first recall some features of axion cosmology in supersymmetric theories~\cite{banksdinegraesser}.  Necessarily in such theories there is 
a modulus which can be thought of as the partner of the axion.  For simplicity, we will assume its mass is of order $m_{3/2}$.  This field starts to
oscillate when $H \sim m_{3/2}$.  The axion starts to oscillate when $m_a \approx H$.  Assuming
that the moduli dominate the energy at this time, we have that the axion energy density is of order $H^2 f_a^2$.  On the other hand,
the modulus energy density is of order $H^2 M_p^2$.  So axions constitute a fraction $ \theta_0^2 f_a^2/M_p^2$ of the energy density.
This is the fraction when the moduli decay (at, say, $10$ MeV.)  In order that axions not dominate the energy density
before temperatures of order $1$ eV, we need
\beq
{\theta_0^2 f_a^2 \over M_p^2} < 10^{-7}\left ( {10 ~{\rm MeV} \over T_r} \right )
\eeq
or
\beq
\theta_0 f_a < 10^{14.5} ~{\rm GeV} \left ( {10 ~{\rm MeV} \over T_r} \right )^{1/2}.
\eeq
If we suppose $f_a$ is given, we have an {\it upper} bound on the reheat temperature, and correspondingly an upper bound
on the mass $m_\phi$.  In particular, for $\theta_0 f_a=10^{14.5}$,
\beq
m_{\phi} \lesssim 100 {\rm ~TeV}.
\eeq

Another upper limit arises from baryogenesis.  Consider first the possibility that baryons are produced in the decays of the $\phi$ particle;
assume that there are $\epsilon$ baryons per decay (independent of the mass of $\phi$.)  In that case, the baryon-to-photon ratio is of order
\begin{align}
\label{baryontophoton}
{n_B \over n_\gamma} =&~\epsilon {T_r \over m_{\phi}}\nonumber\\
 \simeq& ~\epsilon \left( 90 \over \pi^2 g_* \right )^{1/4}\left ({m_{\phi} \over 2\pi M_p} \right )^{1/2}\;.
\end{align}
Alternatively, suppose that the baryons are produced by AD baryogenesis.  We can again parameterize this process in terms of the number of baryons per
modulus, $\epsilon$, and we can again write the baryon-to-photon ratio as in Eq.~(\ref{baryontophoton}).
We can understand the parameter $\epsilon$ more microscopically in such a framework by assuming that the baryon number is generated along
a flat direction described by a pseudomodulus $\Phi$.  We suppose that this field has a mass of order $m_{3/2}$, and that the flat direction
is raised by the appearance in the superpotential of an operator:
\beq
W_B = {1 \over M_p^n} \Phi^{n+3}.
\eeq
We also suppose that in the early universe, there is a term in the $\Phi$ {\it potential}, $-H^2 \vert \phi \vert^2$.  As a result, when $H \approx m_{3/2}$,
the $\Phi$ field begins to oscillate.  Its amplitude is of order
\beq
\Phi^{n+1} \approx M_p^n m_{3/2}\;.
\eeq
Correspondingly, assuming that $W_B$ is baryon-number violating and possesses phase $\delta$,
the baryon number density is:
\beq
n_B \approx m_{3/2} \left ( {m_{3/2}  M_p^n} \right )^{2 \over n+1} \tan(\delta).
\eeq
The number density of moduli at this time is of order $m_{3/2} M_p^2$ (including all moduli, so that $H$ is of order $m_{3/2}$), so
\beq
{n_B \over n_{\gamma}} \approx \left ( {m_{3/2} \over M_p} \right )^{2 \over n+1} \tan(\delta).
\eeq

\section{More Strongly Interacting Moduli; Axions Without Moduli}
\label{strongmoduli}
In string theories, it is often true that at points (or on subspaces more generally) of the moduli space, there are light particles.
At these points, the moduli interactions with themselves may be stronger than expected if they were simply described
by Planck scale local operators.  The modulus lifetime can be much shorter and reheating temperatures higher.  Indeed,
as these are typically points of enhanced symmetry, it is possible that the universe simply finds itself at such a point
as inflation ends, and there is no moduli problem at all~\cite{dinemoduli}.  From the point of view of cosmology, this is similar to the ``no modulus" case we have discussed.

More generally, one can wonder about our use of the Planck scale, as opposed to, say, a scale suppressed by
powers of $g^2$.  As we have just seen, if there is just one modulus, with a logarithmic~\Kahler potential, everything scales with $M_p$.
With more moduli, different scalings are possible, as we saw in our discussion of gaugino masses in the previous section.  Even
with a single field, if one permits more general~\Kahler potential, there are other possibilities.  Still, we view our estimates
of lifetimes and masses as representing a ``typical" behavior away from possible enhanced symmetry points.

To illustrate possible behaviors, suppose that at the enhanced symmetry point, the theory exhibits a {\it linearly} realized symmetry,
under which the modulus (and other fields of the theory) transform by a phase.  The modulus, $X$, might couple to messenger fields,
as in gauge mediation, and other fields, so as to lead to a small breaking of the symmetry.  The low energy theory would contain
operators suppressed by powers of $1/\langle X \rangle$, rather than $1/M_p$\cite{nemeschansky,dinemoduli,lindedudas}.  

We can also contemplate axions which are not parts of moduli fields, according to our definition,
but rather light fields with comparatively flat potentials, perhaps due to a discrete symmetry.  These might resolve
the strong CP problem in theories without moduli or with only supersymmetric moduli, but they must satisfy certain stringent requirements.
These fields could also play a role in the transmission of supersymmetry breaking.
Consider a model~\cite{banksdinegraesser} with a field $\Phi$ coupling to a pair of vectorlike messenger fields and another gauge singlet $S'$,
\begin{align}
W\supset W_0+\Phi\bar{Q}Q+\frac{1}{M_p^{n}}\Phi^{n+2}S'\;
\end{align}
for some integer $n$. We can assume the model respects a discrete
symmetry that accounts for this structure and forbids a linear term in the~\Kahler potential. If $\Phi$ obtains a negative mass from SUSY-breaking $\sim -m_{3/2}^2$, the global symmetry is spontaneously broken, generating a vev $\langle \Phi\rangle\sim (m_{3/2}M_p^n)^{\frac{1}{n+1}}$. The $F$-term is then of order $F_\Phi\sim m_{3/2}\langle \Phi\rangle$ and produces gauge-mediated contributions to the soft masses in the visible sector when the $Q,\bar{Q}$ multiplets are integrated out. At 1-loop, there is a $\Phi W_\alpha^2$ coupling of order $(16\pi^2 \langle \Phi \rangle)^{-1}$.  This  leads to gauge-mediated gaugino masses loop-suppressed relative to $m_{3/2}$, as in the anomaly-mediated contribution.  The 2-loop gauge-mediated scalar masses are of order the squared gaugino masses, and suppressed relative to supergravity contributions. Consequently it is possible for the visible spectrum to remain hierarchical, again bearing similarity to the ``no modulus" case.  

This model has an approximate $U(1)$ global symmetry which can play the role of the Peccei-Quinn symmetry.  It is crucial that this
be a very good symmetry; as is well known, this requires that the underlying discrete symmetry be quite large (e.g. $Z_{12}$).
In models without generic Planck-suppressed moduli, such a structure is necessary to implement the Peccei-Quinn mechanism. The cosmological moduli problem is avoided, and the structure of the visible soft masses is model-dependent.

\section{Conclusions}
\label{conclusions}

While the arguments for TeV scale supersymmetry have long seemed compelling, for some time there have been other reasons to contemplate the possibility
that if supersymmetry plays a role in low energy physics, the scale of supersymmetry breaking might be $10$s of TeV or higher.
The question of cosmological moduli has been among the most troubling of these.  In this paper, we have seen that the presence or absence
of moduli is a controlling consideration for supersymmetry phenomenology.  If there are no moduli, a spectrum with gaugino masses 
smaller by a loop factor than scalar masses seems likely, and WIMP dark matter is produced by conventional thermal processes.
On the other hand, the Peccei-Quinn solution to the strong CP problem is not easily embedded into a UV framework.  In this case, needless to say, moduli
cannot provide an explanation for an unexpectedly high scale of supersymmetry breaking.  In such a picture, the LHC might
find evidence for supersymmetry along the lines discussed in Ref. \cite{AGKWZ}.

If there are only supersymmetric moduli, a split spectrum is again likely, but anomaly mediated contributions to gaugino masses only
dominate for extremely heavy moduli.  If WIMPs are the dark matter, they must be produced in moduli decays or afterwards.  Either requires
a high mass scale.  Again, the Peccei-Quinn solution to the strong CP problem cannot be provided by moduli, and the moduli do not provide an
explanation of any particular scale of supersymmetry breaking.   Avoiding overproduction of gravitinos places significant (but plausible)
constraints on the microscopic theory.

Finally, in the case of non-supersymmetric moduli, a hierarchical or split spectrum is not generic.  If the theory contains a stable
LSP, it is typically overproduced unless the LSP is heavier than the moduli.  This, in turn, implies that the dark matter is likely to be in some form
other than WIMPs.  Assuming axion dark matter, for a fixed axion decay constant, there is an {\it upper} bound on the modulus mass
and correspondingly on the scale of supersymmetry breaking.  Such scenarios point to a supersymmetry breaking scale that is high
compared to the TeV scale, but not arbitrarily high.  

This latter picture suggests that the LHC, at least at $14$ TeV, will not discover evidence for supersymmetry, and that direct and indirect
detection experiments will not find evidence for dark matter.  On the other hand, the next generation of charged lepton flavor violation experiments will permit a new probe of SUSY scales as high as $\mathcal{O}(150)$ TeV~\cite{McKeen:2013dma,Abrams:2012er}.   Such experiments might point to a particular energy
scale. Eventually, a very high energy hadron collider may be able to probe mass thresholds above $10$ TeV directly and perhaps permit the study of supersymmetry breaking at the high scales contemplated here.

\vspace{1cm}

\noindent
{\bf Acknowledgements:}  We thank Nima Arkani-Hamed for conversations about a number of the issues discussed here, and Matt Reece for pointing out an error in the first version of Fig.~\ref{fig:nlsp1}. 
This work supported in part by the U.S.
Department of Energy.

\newpage

\appendix
\section{Supersymmetric and Non-Supersymmetric Decay Amplitudes I}

In section \ref{susydecayrates} (also Appendix B), we showed that in a supersymmetric theory, the decay rates for moduli decays to pairs of particles
are identical to those to their supersymmetric partners.  In this section, we illustrate in detail how this works, in a way
which indicates that the branching ratios are comparable for non-supersymmetric moduli.
We reanalyze each of the specific dimension-5 couplings listed in~\cite{moroirandall}.

Consider first the decay rate into the gauge multiplet. The coupling
\beq
-{A\over 4 M_p} \phi W_\alpha^2
\eeq
generates couplings to gauge bosons and gauginos.  These include: 
\beq
{A \over M_p} \left ( -{1 \over 4}\phi F_{\mu \nu}^2 + {1 \over 4}  F \tilde F + i \phi \lambda \sigma^\mu D_\mu \lambda^* + {1 \over 4} F_\phi \lambda \lambda + {\rm c.c.} 
\right )
\label{phil}
\eeq

In~\cite{moroirandall}, it was noted that the derivative coupling in Eq.~(\ref{phil}) to gauginos is suppressed if the gaugino mass is small.  This can be understood
by a helicity argument, or by using the gaugino equation of motion.  But the term involving the auxiliary field was not considered, and it leads
to a non-negligible coupling of the modulus to the gauginos, even if the
expectation value of the auxiliary field vanishes, as discussed in \cite{nakamurayamaguchi}.  This is in fact what happens in the supersymmetric
case.  Considering, first, global supersymmetry; for a massive field, $F_\phi = m_\phi \phi$, and one has a Yukawa coupling of the modulus
to gauginos, with strength $m_\phi/M_p$.  In the supergravity case, with approximate supersymmetry, the same is readily shown to hold.
If supersymmetry is broken
in supergravity, even if $\phi$ does not appear in the superpotential, writing $\phi = \phi_0 + \delta \phi$,
where $\phi_0$ is the $\phi$ expectation value, one has
\beq
F_\phi = \left ( {\partial^2 W \over \partial^2 \phi} + {\partial^2 K \over \partial^2 \phi} W + \dots \right ) \delta \phi K_2.
\eeq
For light moduli, the second term is typically of order $m_{3/2}$.  This yields the coupling 
\beq
{A \over M_p} e^{K/2} W K_2 {K_2}^{-1}  \delta \phi \lambda \lambda + {\rm c.c.}
\eeq
Scaling $\delta \phi$ so it has canonical kinetic term, and using the relation between $m_{3/2}$ and $W$, this coupling is then
\beq
{ A \over M_p} m_{3/2} {g^2 \over \sqrt{K_2}}  \delta \phi \lambda \lambda + {\rm c.c.}
\eeq
There is no parametric suppression of the branching ratio for light gauginos; of course, once the mass is close to the modulus mass, there
will be phase space suppression.

A similar phenomenon occurs with the other dimension five operators (again as expected from the supersymmetric case).
Consider next the operator
\beq
{B \over M_p} \phi^* H_U H_D + {\rm c.c.}
\eeq
In \cite{moroirandall}, it is stated that this operator does not lead to decay to Higgsinos.  But again this neglects the coupling to the auxiliary component of
$\phi$,
\beq
{B \over M_p} F_\phi^* \psi_{H_U} \psi_{H_D}.
\eeq
Again, in the supersymmetric case, $F_\phi = m_\phi \phi$.
In the non-supersymmetric case, $F_\phi$ includes a term $K_2 W \delta \phi$.
This leads to a coupling, after rescalings (assuming canonical kinetic terms for the Higgs fields)
\beq
K_2^{-1/2} \delta \phi \psi_{H_U} \psi_{H_D}. 
\eeq
If all of the Higgs scalars are lighter than the modulus, then the decay to these fields has the same parametric form. But even if only the lighter Higgs channel
is available, one obtains a similar result, from the coupling:
\beq
{B \over M_p} F_\phi^* (F_{H_U} {H_D} + F_{H_D} H_U) + {\rm c.c.}
\eeq
So again, there is no parametric suppression of the decays to Higgsinos relative to Higgs scalars.

Finally, there are operators of the type: 
\beq
{C \over M_p} \phi Q Q^*
\eeq
The authors of~\cite{moroirandall} note that the decays to light sfermions are suppressed.  After an
integration by parts there is a component operator of the form
\beq
{C \over M_p}  \phi Q(\partial^2 Q^*) + \dots\;,
\eeq
which gives an amplitude proportional to $m_Q^2$. The contribution to the rate is suppressed in the case of supersymmetric moduli. Including also the various auxiliary fields,
\beq
{C \over M_p} ( F_\phi F_Q^* Q + F_Q F_Q^* \phi + \dots)\;.
\eeq
If $Q$ is massless, the decay amplitudes to either fermion or bose pairs again vanish to leading order.  If $Q$ is massive and supersymmetric,
then the couplings $F_\phi Q F_Q^*$ and $\phi F_Q F_Q^*$ contribute to the decay amplitudes, as do the second derivative terms, leading to the expected
equality of decay rates. In fact in this case the leading term in the amplitudes is proportional to $m_Q$, so the rate is suppressed only by two powers instead of four; however, generally the supersymmetric masses are extremely small compared to $m_\phi$. If supersymmetry is broken and the moduli are non-supersymmetric, both the $F$-terms ($F_\phi \approx K_2 W \delta \phi, F_Q \approx Q W$) and the derivative terms give unsuppressed contributions to the decay rates into bosons governed by $(m_Q/m_\phi)^4$.

To summarize, this class of operators generally leads to suppressed decays to ordinary fermions. However, in general, the decays to sfermions are unsuppressed if $m_{\phi}\approx m_{3/2}$, so the operator is certainly problematic with regard to overclosure.

\section{Supersymmetric Decay Amplitudes II}
In this appendix we sketch for illustration a more primitive method to find relations unbroken supersymmetry implies between various decay rates and cross sections. We consider a simple example, the two-body decay of a heavy singlet scalar $\Phi_1$ into two lighter charged scalars $\tilde{q}\bar{\tilde{q}}$ or fermions $q\bar{q}$. The argument is fundamentally equivalent to the Ward identity-LSZ approach given previously\footnote{We note that both arguments are subject to the usual limitation that an unstable state cannot be made asymptotic, so manipulations that treat them as such are only valid in the spirit of the optical theorem and up to corrections of order $\Gamma/M$.}, but does not use field theory.

Consider the sum of the squared matrix elements
\begin{align}
\int d\Omega \sum_i \mathcal{|M|}_i^2=\int d\Omega\sum_i|{}_i\langle q\bar{q}|\frac{1}{2}\epsilon^{\alpha\beta} Q^\dagger_\alpha Q^\dagger_\beta e^{-iHt}|\Phi_1\rangle|^2\;.
\end{align}
Here $Q$ is the Weyl spinor of SUSY generators, $\Phi_1$ is the lowest state (annihilated by both $Q_\alpha$) in a chiral multiplet of mass $M$, and $q$,$\bar{q}$ are the spin-$1/2$ states of two additional $CPT$-conjugate chiral multiplets of mass $m<M/2$. The states $|q\bar{q}\rangle_i$ are the two-fermion states
\begin{align}
|q\bar{q}\rangle_1\equiv|q^{\uparrow}(p)\rangle|\bar{q}^{\downarrow}(-p)\rangle\;,\;\;\;\;\;|q\bar{q}\rangle_2\equiv|q^{\downarrow}(p)\rangle|\bar{q}^{\uparrow}(-p)\rangle\;
\end{align}
and the integration is over solid angle for the outgoing momenta $p$. Acting on the right, the SUSY generators raise $|\Phi_1\rangle$ to $|\Phi_2\rangle$, the highest state (annihilated by both $Q_\alpha^\dagger$) in the $M$ multiplet,
\begin{align}
\mathcal{|M|}_i^2=4M^2|{}_i\langle q\bar{q} |e^{-iHt}|\Phi_2\rangle|^2\;.
\end{align}
Inserting a factor of $(CPT)^2$, 
\begin{align}
\int d\Omega\sum_i|{}_i\langle q\bar{q}|e^{-iHt}(CPT)(CPT)|\Phi_2\rangle|^2=\int d\Omega\sum_i|{}_{i}\langle q\bar{q} |e^{-iHt}|\Phi_1\rangle|^2\;,
\label{rightaction}
\end{align}
because $CPT$ flips the sign of $p$ in the two-fermion states and maps $\Phi_2$ to $\Phi_1$. 

To act on the left with the $Q$, we first decompose the generators into 
\begin{align}
Q_\alpha=Q^{(1)}_\alpha+Q^{(2)}_\alpha(-1)^{F^{(1)}}\;,
\end{align}
where the superscripts denote the one-particle subspaces on which the generators act, and $F^{(1)}$ is the fermion number operator on the first-particle space. The SUSY generators must be moved past the Lorentz generators that boost the fermion momenta to $p$ and $-p$, 
\begin{align}
|q^{\uparrow}(p)\rangle|\bar{q}^{\downarrow}(-p)\rangle=U_{p}^{(1)}{U_{p}^{(2)}}^\dagger|q^{\uparrow}(0)\rangle|\bar{q}^{\downarrow}(0)\rangle\;,
\end{align}
where the axis of spin quantization is assumed for simplicity to lie parallel to $p$ for each $p$. 
$U$s and $Q$s acting on different particle subspaces commute. In terms of spinor components, $U$s and $Q$s acting on the same subspaces obey
\begin{align}
U_pQ^\dagger&=(\Lambda_\frac{1}{2}Q^\dagger)U_p=\left(\begin{array}{c}AQ^\dagger_1\\ A^{-1}Q^\dagger_2\end{array}\right)U_p\;,\nonumber\\
U_p^\dagger Q^\dagger&=(\Lambda_\frac{1}{2}^{-1}Q^\dagger)U_p^\dagger=\left(\begin{array}{c}A^{-1}Q^\dagger_1\\ AQ^\dagger_2\end{array}\right)U_p^\dagger\;,\nonumber\\
A=\sqrt{\gamma(1-\beta)}&\;,\;\;\;\gamma=E/m=M/2m\;,\;\;\;\;\beta=p/E\;.
\end{align}
After some algebra, we obtain
\begin{align}
\int d\Omega\sum_i\mathcal{|M|}_i^2=4m^2(A^4+A^{-4}) \int d\Omega~|\langle \tilde{q}_1\bar{\tilde{q}}_1|e^{-iHt}|\Phi_1\rangle|^2\;.
\label{leftaction}
\end{align}
where $\tilde{q}_1$, $\bar{\tilde{q}}_1$ are the lowest scalar states in the $m$ multiplets and carry momentum $p$ and $-p$.  Equating~(\ref{rightaction}) and~(\ref{leftaction}) and reducing the prefactor, we find
\begin{align}
\frac{1}{2}(1+\beta^2) \int d\Omega~|\langle \tilde{q}_1\bar{\tilde{q}}_1|e^{-iHt}|\Phi_1\rangle|^2=\int d\Omega\sum_i|{}_i\langle q\bar{q} |e^{-iHt}|\Phi_1\rangle|^2\;,
\end{align}
which relates the partial widths of the $\Phi_1$ particle into scalar and fermionic final states. Note that the kinematic factor goes to 1 in the massless limit. Other relations can be derived similarly.

\newpage

\bibliographystyle{unsrt}
\bibliography{dinerefs}{}

\end{document}